\documentstyle[prl,twocolumn,aps,epsf]{revtex}
\draft
\begin{document}

\title{Secondary instabilities form a codimension--2 point accompanied by
       a homoclinic bifurcation}

\author{S. Rudroff$^1$, H. Zhao$^2$, L. Kramer$^2$, and I. Rehberg$^1$}
\address{$^1$ Institut f\"ur Experimentelle Physik,
         Otto-von-Guericke-Universit\"at, D-3916 Magdeburg, Germany \\
         $^2$ Physikalisches Institut, Universit{\"a}t Bayreuth,
          D-95440 Bayreuth, Germany}

\date{\today}
\maketitle

\begin{abstract}

A study of secondary instabilities in
ac--driven electroconvection of a planarly aligned nematic liquid crystal
is presented. At low frequencies one has a transition from normal rolls to
a zig--zag pattern and at high frequencies a direct transition from normal to
abnormal rolls. The crossover defines a codimension--2 point.
This point is also the origin of a line of homoclinic bifurcations, which
enable the transition from zig--zags to abnormal rolls at lower
frequencies.

\end{abstract}


A large degree of universality has been found in primary instabilities
leading to periodic patterns in spatially extended systems\cite{CrHo93}.
A full classification was achieved. For example, in axially anisotropic
quasi--2D systems undergoing a steady supercritical bifurcation one can
have at threshold either normal rolls (NRs), which orient according to the
preferred direction, or oblique rolls, which occur in two
symmetry--degenerate variants and may superpose to give rectangles
\cite{KrPe96}. 

A full classification of the secondary instabilities that may destabilize
the primary patterns appears rather prohibitive, except in the quasi--1D
case \cite{CoIo90}. For Rayleigh-B\'enard convection in simple fluids, the
prime example for isotropic quasi--2D pattern formation, the relevant
secondary instabilities generate the celebrated Busse balloon and have
been analysed in detail \cite{busse}. By contrast, in the best studied
axially anisotropic system, namely ac--driven electroconvection (EC) in
planarly aligned nematic liquid crystal layers, 
the secondary instabilities arising when the main
control
parameter (the amplitude of the voltage) is increased have only recently
been understood starting from the hydrodynamic equations \cite{PlDe97}. It
was found that in the range where one has NRs at threshold the well--known
long--wave zig--zag (ZZ, or undulatory, or wavy) instability occurs only
for frequencies $f$ (the second control parameter) below some value
$f_{AR}$.

For $f>f_{AR}$
one should have a {\em homogeneous}
instability. The nematic director, which near the primary threshold lies
within the plane made up of the NR wave vector ($\parallel {\vec x}$) and
the layer normal ($\parallel {\vec z}$), rotates out of this plane in one
of two symmetry--equivalent directions. Since the director is anchored at
the upper and lower plate along ${\vec x}$, this involves a twist
deformation of the director field, and one may speak of a twist mode. The
resulting roll pattern was termed 'abnormal rolls' (ARs), following
similar
observations in EC with homeotropic anchoring \cite{RiBu94}.
Since homeotropic systems can be rotationally invariant or nearly so
(in the presence of a weak planar magnetic field) the AR--transition
can occur at or near threshold, allowing for a description of the
scenario in terms of coupled Ginzburg--Landau equations \cite{RoHe96}.
However, one of the most interesting effects elucidated in Ref.
\cite{PlDe97}, namely the restabilization of ARs above the
ZZ--instability, is hardly expected under these conditions \cite{RoHe96}.

We here present a study of the predicted crossover from the ZZ-- to the
AR--transition in planar EC. Besides identifying experimentally this new
type of
codimension--2 (C2) point, we unravel the behavior in its neighborhood in
the voltage--frequency plane experimentally and by a normal--form
analysis involving two relevant modes. 
We show that at this C2--point a hitherto unsuspected line
originates, which limits from above the regime where one may have
ZZ--patterns. At this line one has an unusual type of homoclinic
bifurcation from ZZs to ARs. A particularly interesting feature is the
hysteresis found when this line is crossed from above. Then ARs persist
down to a well--defined stability limit. 
The normal form shows that this class of 
two--mode scenario is of general importance, e.g. for certain types
of line defects.

We perform experiments on EC in the conduction
regime of the liquid crystal 4-methoxy\-benzylidene\--4'-n-butyl\-aniline
(MBBA) at a temperature of $15^oC$. We use a channel geometry that is only
moderately extended in the x--direction, namely a capacitor, where a
$25\times 10\,mm$ ITO--coated glass plate faces another glass plate with a
$L_x=315\,\mu m$ wide and $L_y\approx 10000\,\mu m$ stripe of ITO
created by an etching process \cite{HiSh94,RuRe97}. The distance
$d=24\,\mu m$ between the two plates scales the wavelength of the
convection pattern, which arises when the amplitude $V$ of the driving
ac--voltage passes a certain critical value $V_c$. A mechanical treatment
of the surfaces (rubbing) fixes the planar orientation of the director 
almost perpendicular to the y--direction, which is defined by the lateral
boundaries of the channel. The patterns are visualized by the standard
shadowgraph technique \cite{ReHo91} and detected using a microscope and a
video--camera. 

The left--hand side of Fig.\,1 shows the NRs in the low
frequency regime. They arise from the unstructured ground state via a
primary instability, 
and destabilize at a higher voltage due to a
secondary ZZ--instability. The resulting pattern is shown in the middle
part. At an even higher voltage, ARs appear as shown in the right--hand
side of Fig.\,1. They look very similar to NRs---a 
special
optical technique is needed  in order to measure their underlying twist
\cite{GrKl95}. 
This image represents the first direct experimental
observation of the ZZ--AR transition---presumably supported by the lateral
boundaries---in EC in a planarly aligned liquid
crystal. 
\begin{figure}[h]
  \centerline{\epsfxsize=70mm\epsfbox{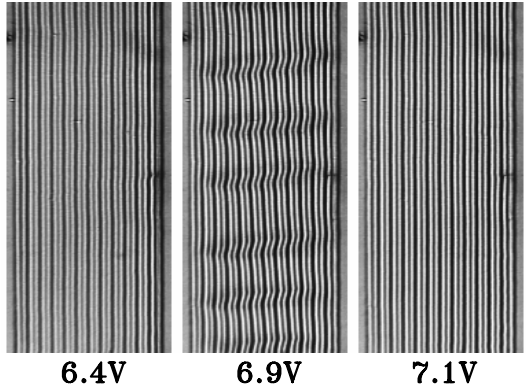}\\[4mm]}
  \caption {\label{figure1.eps}
  Convection patterns in a $315\,\mu m \times
  1020\,\mu m$ part of the channel at 3 different ac--voltages at 43\,Hz.}
\end{figure}

The procedure to obtain the five different relevant transitions
in this scenario is illustrated in Fig.\,2. The measurement of the primary
instability is
presented in the lower part. Here the light intensity modulation
along the $x$--axis was analyzed by means of Fourier analysis.
The intensity of the peak corresponding to the critical wave number
is shown as a function of the driving voltage. This number is
supposed to increase linearly with the distance from the threshold
voltage $V_c$. The linear extrapolation of the data to zero determines
the onset of convection. In our finite channel this
bifurcation is imperfect \cite{HiSh94,RuRe97}. We have measured the
intensity for increasing (open squares) and decreasing (solid circles)
voltage:
This transition has no hysteresis.

The angle of the ZZs
with respect to the boundaries was measured by means of a correlation
technique, at
fixed position where this angle turned out to be maximal. This
order parameter is shown in the middle row of Fig.\,2. It is obvious that
ZZs appear at a driving frequency of 43\,Hz, but not at 80\,Hz. The onset of ZZs at 43\,Hz occurs at 6.55\,V, as
measured by means of a square--root extrapolation. The transition is
imperfect and has no hysteresis. The transition from ZZs to ARs was
determined with a threshold criterion (dotted line). It has a large
hysteresis. The two points measured at the transition from ARs to ZZs are
transients, which cannot be avoided within the pacing of 1 minute used
here.

In order to determine the onset of ARs we have measured the twist angle
by a special optical setup \cite{GrKl95}, which analyzes the ellipticity
of the transmitted light and will be described in detail elsewhere
\cite{RuRe98b}. The results are shown in the upper column of Fig.\,2.
The transition to ARs is easy to identify at the driving frequency of
80\,Hz. An (imperfect) supercritical pitchfork bifurcation is
indicated at a voltage of 7.2\,V: The first
direct experimental demonstration of the transition from NRs to ARs
in EC.

At the driving frequency of 43\,Hz the situation is more complicated. Here
the ZZ--pattern mediates between the regime of NRs and ARs. The ZZs are
accompanied by a twist. Thus the hysteresis in the ZZ--pattern manifests
itself also in a hysteresis of the measured twist angle. The bifurcation
from NRs to ARs can now be identified only indirectly, because both
patterns are unstable with respect to ZZs in the vicinity of this
bifurcation point. We thus measure these points by an extrapolation from
the data obtained in the regime of ARs, assuming an imperfect bifurcation
as indicated by the solid line.
\begin{figure}[h]
  \centerline{\epsfxsize=78mm\epsfbox{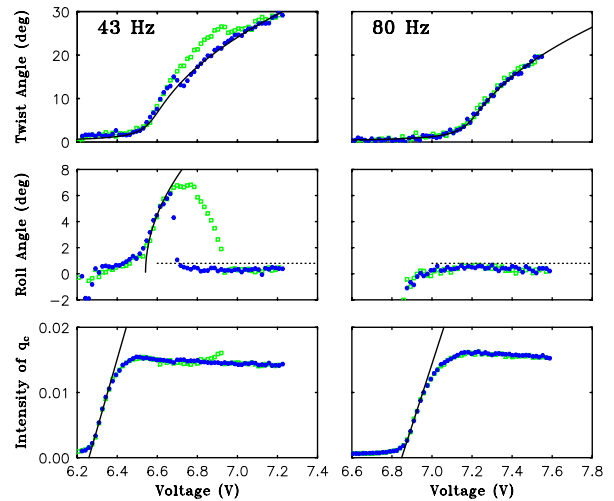}\\[4mm]}
  \caption {\label{figure2.eps}
  Measurements at two frequencies. Open squares (solid circles) correspond
  to
  increasing (decreasing)voltage. We show the
  twist angle (top), the zig-zag angle (middle), and the strength of
  convection (bottom). The solid lines are fits to determine the
  thresholds.}
\end{figure}

Fig.\,3 is a stability diagram, which displays all five
bifurcation lines discussed above. The crosses correspond
to $V_c$. The open squares indicate the
ZZ--instability.
The solid squares (open diamands) mark the transition from ZZs to ARs (ARs
to ZZs) taking place at
increasing (decreasing) voltage. 
The solid circles indicate the bifurcation
from NRs to ARs. In order to simplify a comparison with the theory, the
same data are shown in the upper part of Fig.\,\ref{ecread09.eps} with a
rescaled voltage, namely $\epsilon = V^2/V_c^2 - 1$. Note the similarity
of this measurement with Fig.\,3 of Ref.\cite{PlDe97} (those calculations
were done for a different material). The solid lines are least--square
fits of data points above 67\,Hz to the theory presented below. 
\begin{figure}[h]
  \centerline{\epsfxsize=70mm\epsfbox{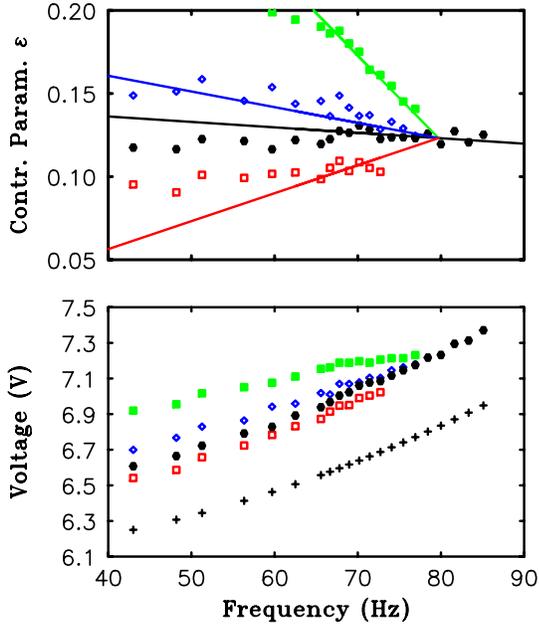}\\[4mm]}
  \caption {\label{figure3.eps}
  Experimental stability diagram. We show the threshold of convection
  (crosses), the  ZZ--instability (open squares), the disappearance of ZZs
  for increasing voltage (solid squares), the destabilization of
  ARs for decreasing voltage (open diamonds), and the
  AR--instability (solid circles).}
\end{figure}

For a theoretical description of the scenario we adopt a phenomenological 
approach based on a Ginzburg-Landau expansion around the bifurcation from 
NRs to ARs in the vicinity of the C2 point. 
Let $\phi$ describe the (real) amplitude of the twist mode, which
plays the role of the order parameter. 
The pre--bifurcation state sustains a spontaneous periodic pattern whose
phase represents a soft (Goldstone) mode which has to be 
included in the description. 
We write the phase as $q {\hat x} + \vec{\nabla} \theta$, such that 
$q {\hat x} + \vec{\nabla} \theta$ is the local wave vector.
Allowing only for variations along $y$ (as is adequate for the channel 
geometry) the Ginzburg-Landau equation for the pitchfork bifurcation from 
NRs to ARs is given by
  \begin{equation}
  \label{AR bif}
  \partial_t \phi = (\mu - g \phi^2)\phi + K \partial_y^2 \phi
  - \gamma (\partial_y \theta + P_a),
  \end{equation}
with positive $g$ and $K$ and (reduced) control parameter $\mu$. 
The last term describes the bias when the rolls are (slightly)
oblique (the local roll angle is $\arctan(\partial_y\theta/q)$). We
allowed for a misalignment between the rubbing direction ($\phi=0$) and
the $x$--axis defined by the channel geometry. $\gamma$ must be chosen
positive, so reorienting the rolls favors rotation of the director
in the opposite sense \cite{PlDe97}. From a fit to the experiment
(Fig.\,2, top right) one finds $\gamma g^{1/2} P_a=0.0011$.
The dynamics of the phase modulation is governed by the equation
  \begin{equation}
  \label{phase diff}
  \partial_t\theta = \partial_y J,\
  J=D (\partial_y \theta + P_a)- (\nu + h \phi^2/3)\phi,
  \end{equation}
with positive $D$ and $h$ (as it turns out).
The first term in $J$ describes ordinary phase diffusion, and the 
second expression represents the coupling to the twist mode. The
nonlinear term is essential, because $\nu$ crosses zero in the region
of interest.

Straight--roll solutions 
(normal or oblique) are characterized by $\phi=\phi_s~(=const.)$ and
$\partial_y \theta=P~(=const.)$ and their stability is derived from
the growth rate $\lambda=-\frac{1}{2}B+\sqrt{\frac{1}{4}B^2 - C}$
of modes $\sim e^{\imath p y}$, where $B=-\mu+3g \phi_s^2 + (K+D)p^2,\
C=[\gamma \nu -D\mu + 3(gD+h\gamma/3) \phi_s^2]p^2+ KDp^4$.
For stability one needs $B>0$ and $C>0$. Thus, for NRs with $\phi_s=0$
and $P_a=0$, there is a homogeneous instability at $\mu=0$, leading
to ARs and a long--wave ZZ--instability at $\mu=\mu_{zz}=S_{zz} \nu$ with
$S_{zz}=\gamma/D$. For negative $\nu$ one first has the
ZZ--instability and eqs.\,(\ref{AR bif},\ref{phase diff}) indeed
describe the observed crossover scenario, as indicated in
Fig.\,\ref{ecread19.eps}, lines AR and ZZ. From the expression
for $C$ one sees that a nonzero $\phi$ suppresses the ZZ--instability.
For negative $\nu$ this effect leads to restabilization of ARs
($\phi_{AR}^2= \mu/g$) above the line $\mu= S_{rs}\nu$ with
$S_{rs}=\frac{S_{zz}}{3 S_{zz}/S_{hb}-2}$, $S_{hb}=-3 g/h$,
see Fig.\,4, line RS.
\begin{figure}[h]
  \centerline{\epsfxsize=70mm\epsfbox{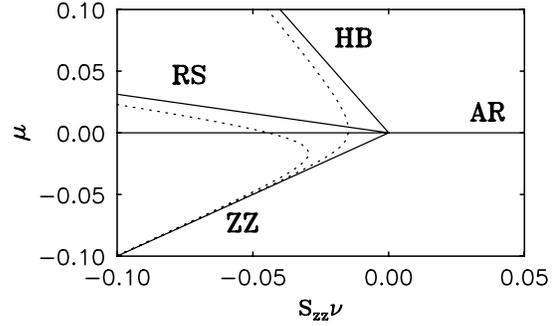}\\[4mm]}
  \caption{\label{figure4.eps}
  Theoretical stability diagram for $S_{zz}/S_{hb}=-2/5$.
  Solid (dashed) lines are for $P_a=0$ ($\gamma g^{1/2} P_a=0.0011$).}
\end{figure}

We now analyse the scenario expected when $\mu$ is varied for
negative $\nu$. For static solutions of (\ref{AR bif},\ref{phase diff})
one has $J = const$. Eliminating $\partial_y \theta$ leads to
  \begin{equation}
  \label{stationary}
  K\partial_y^2 \phi=-(\mu - \mu_{zz})\phi+ g'\phi^3 + S_{zz} J,
  \end{equation}
with $g'=g(1-S_{zz}/S_{hb})$, which is 
integrable. 
Invoking the analogy of a point particle
(coordinate $\phi$, time $y$) one sees that 
the bounded solutions are either constant or periodic. 
The channel geometry requires
that rolls remain 
on the average oriented along $y$, so that
the average of 
$\partial_y \theta$ is zero.
From (\ref{phase diff}) one then has
$J= D P_a- <(\nu + \frac{1}{3}h \phi^2)\phi>$ ($<...>=$ spatial average).

First we look for ZZ--solutions in the case $P_a=0$, where $\phi$
oscillates symmetrically around zero ($J=0$). From
(\ref{stationary}) one gets a one--parameter family of periodic
solutions above the ZZ--line (apart from phase shifts).
In particular there is the homoclinic (or actually heteroclinic) limit
where the solution degenerates to a widely spaced array of domain walls
separating regions where $\phi$ approaches the constant solutions
$\pm \phi_{zz}$ with $\phi_{zz}= \sqrt{(\mu-\mu_{zz})/g'}$.
The undulations observed under increase of the voltage are
(presumably) approximated by this solution. The maximum value of the
roll angle is given by $\arctan(p_{zz}/q)$ with
$p_{zz} =\frac{h}{3 D g'^{3/2}} (\mu-\mu_{zz})^{1/2} (\mu-\mu_{hb})$.
The roll angle (and thereby the undulation) first increases
with $\mu$ and then decreases, reaching zero at the line 
$\mu=\mu_{hb}=S_{hb}\nu$. There $\phi_{zz}$ coincides with
$\phi_{AR}$ and one is left with an array of marginally stable domain
walls separating AR--regions with alternating director twist.
Above this line (and also for positive $\nu$) only moving
domain walls exist stably 
(for details see
\cite{ZaKr98}).

Therefore, when the line HB is reached, the domain walls annihilate
pairwise, and a single--domain AR is established.
From now on ARs persist under changes of the parameters until their
stability limit is reached. The integration constant $J$ maintains
the AR value $-(\nu + \frac{1}{3}h \phi_{AR}^2)\phi_{AR}
\ (=0$ at $\mu_{hb})$.
Thus, on lowering $\mu$, ARs persist down to the line RS where
a discontinuous ZZ--instability occurs, as observed.
The twist angle in the ZZ--state is larger than in the coexisting ARs, 
in agreement with the experiment (Fig.\,2, top left).
The solid lines in Fig.\,3 represent a five-parameter fit of 4 lines
through a common
intersection point (the C2 point) with slopes incorporating the relation
$S_{rs}=\frac{S_{zz}}{3 S_{zz}/S_{hb}-2}$. This relation is in fact
invariant under the linear mapping that connect the control parameters
$\mu$ and $\nu$ of the model with the experimental control parameters
$\epsilon-\epsilon_{AR}$ and $f-f_{AR}$.

The observed asymmetry of the undulations can be accommodated by choosing
$P_a \ne 0$. In Fig.\,4 the corresponding scenario is shown (dashed
lines). The ZZ--instability remains sharp, since translation invariance
along $y$ remains intact. At the homoclinic bifurcation again $J=0$. Now
the length of the shorter arms of the ZZs vanishes there. The selected
wavelength of the observed ZZ undulations presumably results from the
finite width of the channel
\cite{ZaKr98}.


The nature of the hysteresis found is actually quite unique. One
ingredient, typical for pattern--forming systems, is the conservation of
the number of periodic units (here ZZ--undulations) due to pinning at the
boundary (here the short ends). Another ingredient is the unusual
homoclinic bifurcation of ZZ patterns from ARs. We note that the
ZZ--instability comes out naturally, which underscores its importance for
EC. In fact, for many other nematics, like the one used in Ref.
\cite{PlDe97}, the ZZ--instability line joins the primary bifurcation line
at another significant C2--point, which separates the regimes where
oblique and normal rolls appear at threshold.

In summary, two nontrivial qualitative features of the model are in
accordance with the experimental data: A hysteretical transition between
ZZs and ARs vanishes at the C2--point, and the width of this hysteresis
decreases linearily with the distance from this point.

The model should be applicable to other instabilities exhibiting a similar
symmetry. Candidates are line defects and some types of domain walls in
the bend--Fr\'eedricksz distorted state in nematics. Sometimes in these 1D
extended structures the director can escape out of the symmetry plane,
which may mimic the transition from NRs to ARs described by $\phi$. If the
position of the line/wall is not fixed from outside it can be described by
our phase variable $\theta$. In those cases where one has a potential (no
dissipative driving) our model predicts $\gamma \nu < 0$ so that the
ZZ--instability always occurs first, which appears to be consistent with
experiments \cite{bend}. The model is not applicable if the coupling terms
between the two active modes vanish by symmetry, as is the case in
Ising--Bloch--type transitions of domain walls \cite{CoLe90}. We also
remark that
the validity of the model is restricted to the
immediate neighborhood of the C2 point.

We 
thank C. Chevallard, W. Pesch, E. Plaut, A. G. Ro{\ss}berg and
R. Stannarius for discussions and help. 
Support by DFG through
Re588/12 and EU through TMR network FMRX-CT96-0085 is gratefully
acknowledged.

\end{document}